\documentclass[manuscript, screen]{acmart}

\usepackage{algorithmic}
\usepackage{graphicx}
\usepackage{textcomp}
\usepackage{xcolor}

\usepackage{booktabs}
\usepackage{caption}

\usepackage{multirow} 
\usepackage{array}
\usepackage{makecell}
\usepackage{url}
\usepackage[shortlabels]{enumitem}
\usepackage{xspace}

\newcommand{\codellama}{CodeLlama}

\newcommand{\declcomp}{RFCS\xspace}
\newcommand{\funcslice}{HFCS\_m\xspace}
\newcommand{\commslice}{HFCS\_mc\xspace}

\newcommand{\hier}{\textit{HMCS}\xspace}

\usepackage{pifont}
\usepackage{tcolorbox}
\newcommand{\finding}[1]{
\begin{center}
\begin{tcolorbox}[colback=white!15, colframe=black, boxsep=-0.15cm, middle=-0.15cm]
\textbf{\ding{43} Finding}
$\blacktriangleright$
{#1}
$\blacktriangleleft$
\end{tcolorbox}
\end{center}
}
\newcommand{\summary}[1]{
\begin{center}
\begin{tcolorbox}[colback=gray!15, colframe=black, boxsep=-0.15cm, middle=-0.15cm]
\textbf{\ding{46} Summary}
$\blacktriangleright$
{#1}
$\blacktriangleleft$
\end{tcolorbox}
\end{center}
}

\usepackage{etoolbox}
\makeatletter
\patchcmd{\authornote}{\g@addto@macro\addresses{\@authornotemark}}{}{}{}
\makeatother

\usepackage[normalem]{ulem}

\AtBeginDocument{%
  }

\setcopyright{acmlicensed}
\copyrightyear{2018}
\acmYear{2018}
\acmDOI{XXXXXXX.XXXXXXX}

\begin{document}

\title{Commenting Higher-level Code Unit: Full Code, Reduced Code, or Hierarchical Code Summarization}

\author{Weisong Sun}

\email{weisong.sun@ntu.edu.sg}
\orcid{0000-0001-9236-8264}
\affiliation{%
  \institution{Nanyang Technological University}
  \city{Singapore}
  \country{Singapore}
}

\author{Yiran Zhang}     
\email{yiran002@e.ntu.edu.sg}
\orcid{0000-0002-9366-6076}
\affiliation{
  \institution{Nanyang Technological University}
  \city{Singapore}
  \country{Singapore}
}

\author{Jie Zhu}     
\email{1164520422@qq.com}
\orcid{0009-0003-3667-1447}
\affiliation{
  \institution{Yangzhou University}
  \city{Yangzhou}
  \country{China}
}

\author{Zhihui Wang}     
\email{1478852101@qq.com}
\orcid{0009-0004-8554-9277}
\affiliation{
  \institution{Yangzhou University}
  \city{Yangzhou}
  \country{China}
}

\author{Chunrong Fang}
\email{fangchunrong@nju.edu.cn}
\orcid{0000-0002-9930-7111}
\affiliation{
  \institution{State Key Laboratory for Novel Software Technology, Nanjing University}
  \city{Nanjing}
  \country{China}
}

\author{Yonglong Zhang}     
\email{ylzhang@yzu.edu.cn}
\orcid{xxx}
\affiliation{
  \institution{Yangzhou University}
  \city{Yangzhou}
  \country{China}
}

\author{Yebo Feng}
\email{yebo.feng@ntu.edu.sg}
\orcid{0000-0002-7235-2377}
\affiliation{
  \institution{Nanyang Technological University}
  \city{Singapore}
  \country{Singapore}
}

\author{Jiangping Huang}
\email{huangjp@cqupt.edu.cn}
\orcid{0000-0002-0288-6824}
\affiliation{
  \institution{Chongqing University of Posts and Telecommunications}
  \city{Chongqing}
  \country{China}
}

\author{Xingya Wang}
\email{xingyawang@outlook.com}
\orcid{0000-0002-7331-4831}
\affiliation{
  \institution{Nanjing Tech University}
  \city{Nanjing}
  \country{China}
}

\author{Zhi Jin}
\email{zhijin@pku.edu.cn}
\orcid{0000-0001-7300-9215}
\affiliation{
  \institution{School of Computer Science, Wuhan University; Key Laboratory of High Confidence Software Technologies (Peking University), Ministry of Education; School of Computer Science, Peking University}
  \city{Beijing}
  \country{China}
}

\author{Yang Liu}
\email{yangliu@ntu.edu.sg}
\orcid{0000-0001-7300-9215}
\affiliation{
  \institution{Nanyang Technological University}
  \city{Singapore}
  \country{Singapore}
}

\renewcommand{\shortauthors}{W. Sun, Y. Zhang, J. Zhu, Z. Wang, C. Fang, Y. Zhang, Y. Feng, J. Huang, X. Wang, Z. Jin, Y. Liu}

\begin{abstract} 
Commenting code is a crucial activity in software development, as it aids in facilitating future maintenance and updates. To enhance the efficiency of writing comments and reduce developers' workload, the software engineering community has proposed various automated code summarization (ACS) techniques to automatically generate comments (also called summaries) for given code units. 
However, these ACS techniques, including the recent ACS techniques based on large language models (LLMs), primarily focus on generating summaries for code units at the method level. 
There is a significant lack of research on summarizing higher-level code units, such as file-level and module-level code units, despite the fact that summaries of these higher-level code units are highly useful for quickly gaining a macro-level understanding of software components, modules, and architecture. 

To fill this gap, in this paper, we conduct a systematic study on how to use LLMs for commenting higher-level code units, including file level and module level. These higher-level units are significantly larger than method-level ones, which poses challenges in handling long code inputs within LLM constraints and maintaining efficiency. To address these issues, we explore various summarization strategies for ACS of file-level and module-level code units, which can be categorized into three types: full code summarization, reduced code summarization, and hierarchical code summarization. As the name suggests, full code summarization involves inputting all the code of the higher-level code unit into LLMs as context. Reduced code summarization only inputs the key information, excluding details like method bodies. Hierarchical code summarization involves first generating summaries for lower-level code units (e.g., methods for files, files for modules), and then using those summaries to generate summaries for higher-level code units. We investigate three LLMs (including one general-purpose LLM and two specialized code LLMs) to generate summaries based on summarization strategies. The experimental findings from human evaluations suggest that for summarizing file-level code units, using the full code is the most effective approach, with reduced code serving as a cost-efficient alternative. However, for summarizing module-level code units, hierarchical code summarization becomes the most promising strategy. In addition, inspired by the research on method-level ACS, we also investigate using the LLM as an evaluator to evaluate the quality of summaries of higher-level code units. The experimental results demonstrate that the LLM's evaluation results strongly correlate with human evaluations, and most of the quantitative conclusions remain consistent between both LLM and human evaluators.
We hope that our findings will promote the development and application of ACS on higher-level code units.
\end{abstract}

\ccsdesc[500]{Software and its engineering~Documentation}

\keywords{Large Language Model, Higher-level Code Unit Summarization, Automatic Code Summarization, Hierarchical Code Summarization}

\maketitle

\section{Introduction}
\label{sec:introduction}
Code comments (also known as summaries) are crucial for facilitating program code understanding~\cite{1981-Comments-on-Program-Comprehension} and maintenance~\cite{2005-Documentation-Essential-Software-Maintenance}. 
However, high-quality summaries are frequently absent, mismatched, or outdated during program code evolution~\cite{2005-Documentation-Essential-Software-Maintenance, 2023-EACS}. 
Automatic code summarization (ACS), aimed at automatically generating natural language summaries for given code, has consistently been a popular research direction in the field of software engineering (SE)~\cite{2022-Evaluation-Neural-Code-Summarization, 2024-SIDE, 2023-Prompt-CS, 2024-Distilled-GPT-for-Code-Summarization, 2025-LLM4CodeSum}. 

Existing ACS studies mainly focus on function/method-level code summarization. 
These studies have significantly improved the quality of the generated method-level code summaries, providing great assistance to developers in understanding method code. 
However, there is very limited research on summarizing higher-level code units, such as file-level and module-level code. 
Summaries of these higher-level code units are highly valuable for quickly gaining a macro-level understanding of various program functionalities, modules, and architectures. 
In recent years, with the tremendous success of large language models (LLMs) in the field of natural language processing (NLP), an increasing number of SE researchers have adapted them for solving SE tasks, including the ACS task. 
In addition to directly applying general-purpose LLMs (e.g., Llama 3~\cite{2024-Llama-3} and GPT-3.5/4~\cite{2023-GPT-4}), SE researchers have also trained a series of specialized LLMs for code, such as CodeLlama~\cite{2023-CodeLlama} and CodeGemma~\cite{team2024codegemma}.
There are typically two approaches to applying LLMs to specific downstream tasks: 1) supervised fine-tuning on specific downstream task datasets; and 2) instructing the LLMs to complete specific downstream tasks through prompts, including advanced prompt engineering techniques. 
Since the first approach involves model training, which requires significant resources and cost, the second approach has become more favored by users and is gradually becoming a trend. 
Recent efforts on LLM-based ACS~\cite{2022-Few-shot-Training-LLMs-for-Code-Summarization, 2023-Automatic-Code-Summarization-via-ChatGPT, 2024-LLM-Few-Shot-Summarizers-Multi-Intent-Comment-Generation, 2025-LLM4CodeSum} have also confirmed this trend. 
Similarly, most of these LLM-based ACS efforts also focus on method-level code summarization. The summarization capabilities of LLMs for higher-level code units have yet to be explored. 

To fill these gaps, in this paper, we conduct the first study on the summarization of higher-level code units based on LLMs. 
Specifically, we focus on the ACS of two higher-level code units: the file-level ACS, which aims to generate a summary for a given file code, and the module-level ACS, which aims to generate a summary for a given module code. 
Considering that both general-purpose LLMs and specialized code LLMs have limited input context windows, we explore different summarization strategies for these two levels of ACS. 
Specifically, for the file-level ACS, in addition to inputting the full file code (if it does not exceed the model's context window limit), we introduce the following two summarization strategies for comparison.
The first strategy is referred to as file code reduction, which follows the work~\cite{li2023classsum} and removes the implementation details (i.e., method code bodies) of methods and only retains key information in the file code, including class signature, global fields, and method signatures. 
The second strategy is called file hierarchical summarization, which first generates a summary for each code fragment in the file code, and then generates a file summary by further summarizing these fragment summaries. 
We investigate two types of fragments: one treats each method in the file code as a fragment, and the other treats a combination of methods with call relationships within the file code as a fragment. We refer to these combinations of methods as forming a method community, since it is identified by the community algorithm~\cite{blondel2008fast}. 
The experimental results show that for file level, using full code can generate the best summarization result since both of the two strategies may introduce information loss. However, the file code reduction strategy can also generate summaries of reasonable quality at a 74.65\% lower cost.
For the module-level ACS, since a module typically consists of multiple files and contains a larger volume of code, summarization strategies similar to the file-level ACS may be even more necessary. 
Similarly, besides using the full code, we explore two summarization strategies for the module-level ACS: 1) module code reduction, which is similar to file code reduction and removes the implementation details of methods in each file code and only retains key information. 
2) module hierarchical summarization, which first generates a summary for each file code in the module code, and then summarizes these file code summaries to produce a module summary. 

The experimental results indicate that, in contrast to the file level, using the full code at the module level leads to the worst summary quality and higher costs. Hierarchical summarization proves to be the most effective method at this level by receiving the highest scores in human evaluations. Meanwhile, the code reduction strategy offers a cost-effective alternative, yielding better outcomes than the full code summarization at a significantly reduced cost.

It should be noted that the above results are based on large-scale manual evaluations. Compared to evaluating method-level code summaries, where evaluators only need to review short method code, evaluating summaries of higher-level code units is much more time-consuming and labor-intensive, as evaluators need to read through large amounts of file and module code. Furthermore, unlike the more mature research on method-level ACS, higher-level ACS research still lacks automated evaluation methods. Considering that we are the first to investigate higher-level code unit summarization, and to promote the development of this field, we attempt to explore the feasibility of using LLMs themselves as evaluators to assess the quality of higher-level code summaries, inspired by the research on method-level code summarization~\cite{2025-LLM4CodeSum}. 
Specifically, we use GPT-4 to act as a human evaluator and rate the quality of higher-level code summaries on a scale from 1 to 5. 
The experimental results show a high positive correlation between the scores given by the GPT-4 evaluator and those from human evaluators. This suggests that future researchers can use GPT-4 as a substitute for human evaluators to conduct large-scale quality assessments of summaries of higher-level code units.

In summary, we make the following contributions:
\begin{enumerate}

    \item To the best of our knowledge, we are the first to conduct a systematic study on the ACS of higher-level code units based on LLMs. We compare various summarization strategies for file-level and module-level code summarization. Our findings suggest that the full code summarization is the most effective strategy for file-level ACS, while reduced code serves as a cost-efficient alternative. For module-level ACS, hierarchical summarization is the most promising strategy. These findings and insights can benefit future research and practical usage of higher-level code summarization. 

    \item We reveal the feasibility of applying the LLM itself as an evaluator to assess the quality of higher-level code summaries. This helps promote the automation and benchmarking of evaluations in this field.

    \item We make our dataset and source code publicly accessible~\cite{2025-LLM4ModuleSum} to facilitate the replication of our study and its application in extensive contexts.

\end{enumerate}

\section{Background and Related Work}
\label{sec:background_and_related_work}
Code summarization~\cite{steidl2013quality, zhang2022survey}, which is also known as code commenting, is the textual description of the functionality and purpose of special identifiers in computer programs. 
In other words, it uses natural language to explain/summarize the logic and functionality of the source code, making it easier for people to understand the program. It is well known that software maintenance is the most expensive and time-consuming phase in the software lifecycle~\cite{1980-Stability-Measures-for-Software-Maintenance}. High-quality code summary/comment is essential for understanding and maintaining programs~\cite{song2019survey}. For example, it can reduce the time developers need to understand the source code and improve code readability. Unfortunately, with the rapid evolution of software, most code summaries are mismatched, outdated, and missing, requiring continuous improvement and updating~\cite{2018-DeepCom}.

Deep learning models~\cite{2021-CodeT5, feng2020codebert, wan2018improving, wang2020reinforcement, li2023classsum} have spearheaded the latest developments in SE tasks like code summarization. Although these models perform well, they require fine-tuning with large amounts of labeled data. In recent studies~\cite{2023-Automatic-Code-Summarization-via-ChatGPT, nam2024using, 2024-Distilled-GPT-for-Code-Summarization}, LLMs have been applied as one of the primary methods for automated code summarization. Additionally, in several experiments, they have outperformed some deep learning models~\cite{2022-Few-shot-Training-LLMs-for-Code-Summarization, 2024-LLM-in-Statement-level-Code-Summarization}.

In recent research~\cite{2024-LLM-Few-Shot-Summarizers-Multi-Intent-Comment-Generation, 2023-Automatic-Code-Summarization-via-ChatGPT, fried2022incoder, 2022-Few-shot-Training-LLMs-for-Code-Summarization, leclair2020improved, chen2021my}, many scholars have employed various methods to improve the performance of LLMs in code summarization. Fried et al.~\cite{fried2022incoder} propose InCoder, a large generative model for code that performs code infilling and synthesis using bidirectional context. They demonstrate the model's effectiveness through zero-shot evaluation on various code summarization tasks, achieving competitive results. Ahmed et al.~\cite{2022-Few-shot-Training-LLMs-for-Code-Summarization} investigate the effectiveness of few-shot training with LLMs for project-specific code summarization. They show that by using only a few examples, the Codex model with few-shot training can outperform state-of-the-art fine-tuned models, highlighting the potential of leveraging limited project-specific data to enhance code summarization performance.

\begin{figure}[!t]
    \centering
    \includegraphics[width=\linewidth]{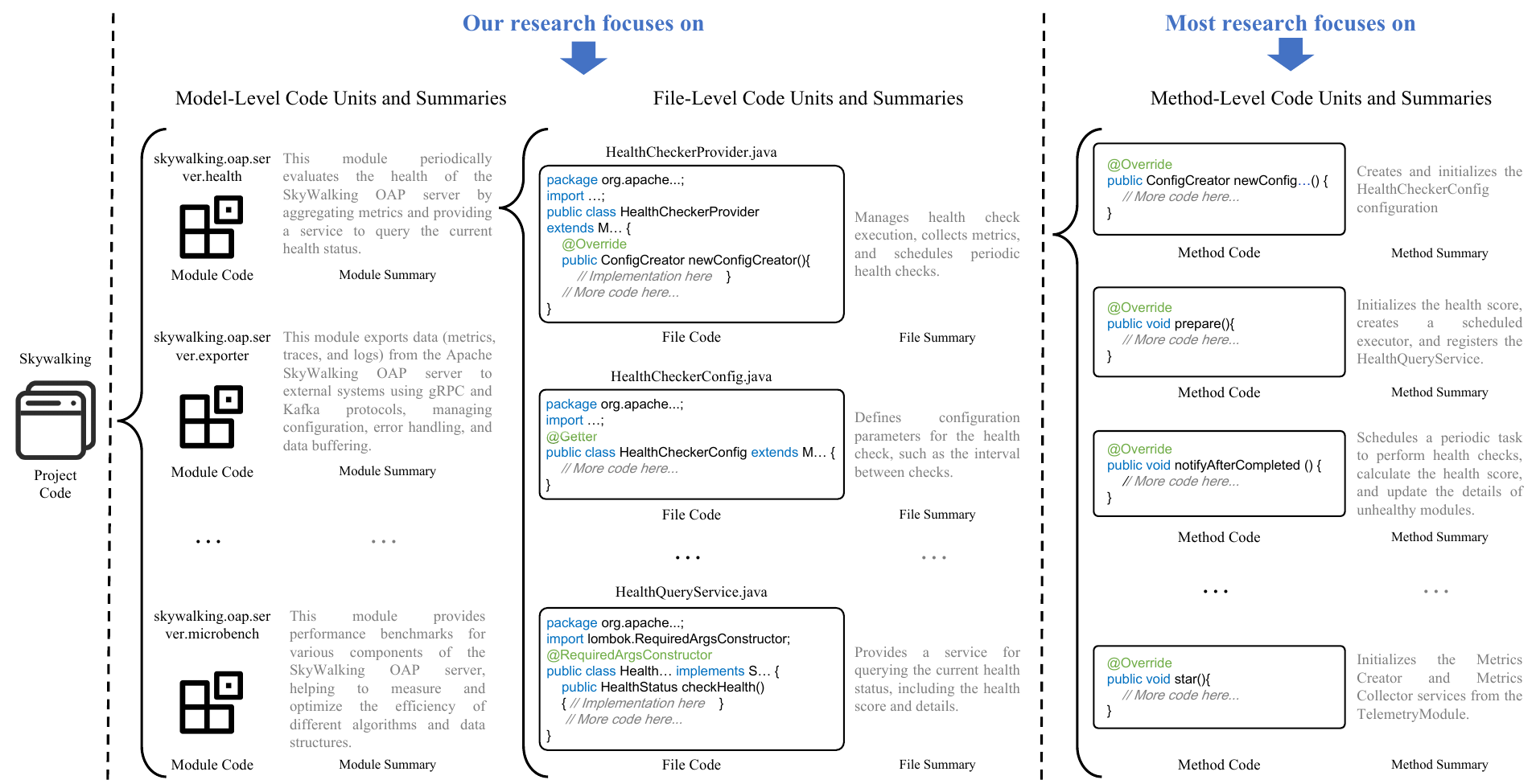}
    \caption{An example of ACS on different level code units.}
    \Description{An example of ACS on different levels of code units.}
    \label{fig:example_of_ACS_on_higher-level_code_unit}
\end{figure}

However, as shown in Fig~\ref{fig:example_of_ACS_on_higher-level_code_unit}, most of these efforts~\cite{mcburney2015automatic,2025-LLM4CodeSum,shi2022we,2023-Automatic-Code-Summarization-via-ChatGPT} concentrate on improving method-level code summarization, while research on more granular code units, file-level code summarization~\cite{wang2024sparsecoder}, and module-level code summarization remains underexplored. SparseCoder~\cite{wang2024sparsecoder} utilizes sparse attention mechanisms for efficient file-level summarization of long code sequences. However, it struggles with context window limitations for module-level summarization and overlooks the interactions and dependencies at a higher level of abstraction. ClassSum~\cite{li2023classsum} is a deep learning model for class-level code summarization in Java. However, its performance drops significantly in cross-project tests. This paper, therefore, aims to enhance the capabilities of LLMs for performing ACS on both file-level and module-level code units, introducing hierarchical summarization strategies to address these challenges.

From Fig.~\ref{fig:example_of_ACS_on_higher-level_code_unit}, it is observed that the summaries of higher-level code units (including file-level and module-level code units) are highly useful for quickly gaining a macro-level understanding of software components, modules, and architecture. For example, developers can quickly understand the functionality of the \texttt{skywalking.oap.server.health} component in the \texttt{SkyWalking} project by reading its short summary: ``\textit{This module periodically evaluates the health of the SkyWalking OAP server by aggregating metrics and providing a service to query the current health status.}'' This is much more efficient than first analyzing multiple code files within the module (e.g., \texttt{HealthCheckerProvider.java}, \texttt{HealthCheckerConfig.java}, \dots, \texttt{HealthQueryService.java}) individually and then trying to grasp the overall functionality of the module. 
In addition, as we all know, software architecture~\cite{garlan1993introduction} is the core of designing complex systems, involving the organization, communication, and functional allocation of components. As systems grow in scale, architecture becomes more critical than algorithms and data structures. A proper architecture enhances system scalability and maintainability, while a poor one can lead to failure. Therefore, understanding and applying software architecture is essential for successful software development. File-level code summaries (e.g., the file summary ``\textit{Manages health check execution, collects metrics, and schedules periodic health checks.}'') help in understanding the specific functionality and logic of individual files (e.g., the file \texttt{HealthCheckerProvider.java}), making it easier to grasp their role within the system. 
Module-level code summaries provide an overall view of a module, revealing relationships and functional collaboration between modules. For example, from the summaries of the two modules \texttt{skywalking.oap.server.health} and \texttt{skywalking.oap.server.exporter}, they collaborate to complete the evaluation and reporting of the quality of the Skywalking OAP server. 
Together, the file-level and module-level summaries offer a comprehensive understanding of the software system, from local details to the broader architecture.

\section{Study Design}
\label{sec:study_desigh}

\subsection{Research Questions}
\label{subsec:research_questions}
\noindent This study aims to answer the following research questions:

\textbf{RQ1: How can high-quality file-level code summaries be effectively generated?} 
When generating file-level code summaries using LLMs, two important issues need to be addressed: 
1) The context window limit of LLMs. Different LLMs have varying capacities for handling input context, i.e., different context window limits. Therefore, we must consider how to generate file-level code summaries when the number of tokens in a code file exceeds the context window limit of the LLM. 
2) The budget constraint for using LLMs. It is well-known that commercial LLMs (e.g., GPT-3.5 and GPT-4) charge based on the number of input tokens. When a code file contains a large number of tokens, even if it does not exceed the model's context window limit, we may still face budget constraints. In this case, we must consider how to obtain high-quality file-level code summaries under a limited budget.
Therefore, in this RQ, we aim to answer two sub-questions: 
1) \textit{Is it necessary to input the entire file content into the LLM when generating file-level code summaries?} 
2) \textit{When the number of tokens in a code file exceeds the context window limit of the LLM or exceeds the available budget, what are the alternative strategies for generating file-level code summaries?} 

\textbf{RQ2: How can high-quality module-level code summaries be effectively generated?} 
Since a software module typically consists of multiple code files, generating module-level code summaries must address the same two issues encountered in file-level code summarization. Thus, similarly, in this RQ, we aim to answer the following two sub-questions: 
1) \textit{Is it necessary to input the entire module content into the LLM when generating module-level code summaries?} 
2) \textit{When the number of tokens in a module exceeds the context window limit of the LLM or exceeds the available budget, what are the alternative strategies for generating module-level code summaries?} 

\textbf{RQ3: How can the quality of file-level and module-level code summaries be automatically evaluated?} 
Although existing research on method-level code summarization offers various automated evaluation methods, most of them rely on datasets containing ground-truth summaries. However, in real-world code summarization applications, such ground-truth summaries are often unavailable. 
Furthermore, although evaluating the summary quality by humans may be the most accurate, it is impractical when dealing with a large amount of code related to the summary (e.g., a module summary may be related to hundreds of code files). Manually reading through extensive code to assess the quality of corresponding summaries is not feasible. 
In this RQ, we aim to reveal whether the LLM itself can be used as an evaluator for evaluating the quality of generated file-level and module-level code summaries.

\subsection{Experimental LLMs}
\label{subsec:experimental_LLMs}

\noindent We select three LLMs as experimental representatives, including CodeLlama, CodeGemma, and GPT-4.
We select CodeLlama and CodeGemma as the experimental models due to their widespread use as open-source code generation models, making them highly representative in the field. Additionally, we include GPT-4 as a representative closed-source model, as it has set the standard for high-performance capabilities in NLP and code generation tasks. GPT-4 serves as a prime example of a closed-source solution that demonstrates the potential of advanced models in the field, offering valuable insights for comparing and contrasting with open-source models like CodeLlama and CodeGemma.

\textbf{\codellama{}.} Code Llama~\cite{2023-CodeLlama} is a family of LLMs for code based on Llama 2~\cite{2023-Llama-2}.  
It provides multiple flavors to cover a wide range of applications: foundation models, Python specializations (Code Llama-Python), and instruction-following models (Code Llama-Instruct) with 7B, 13B, and 34B parameters.  
Our study utilizes Code Llama-Instruct-7B. The recommended context window size is 16K tokens, but it supports up to 100K tokens as the maximum input.

\textbf{CodeGemma.} CodeGemma~\cite{team2024codegemma} is a collection of powerful, lightweight models that can perform a variety of coding tasks like fill-in-the-middle code completion, code generation, natural language understanding, mathematical reasoning, and instruction following. Our study utilizes Codegemma-7B. The Context window size is 8k tokens.

The choice of the 7B versions was primarily driven by resource constraints, as these models have demonstrated strong performance in numerous studies~\cite{2024-LLM-A-Suervey} while requiring fewer computational resources compared to larger versions. This choice strikes a balance between maintaining high performance and managing computational demands effectively.

\textbf{GPT-4.} GPT-4~\cite{2022-OpenAI} is an LLM provided by OpenAI. It is trained with massive texts and codes. OpenAI has not disclosed the specific parameter scale of GPT-3.5 and GPT-4. Our study uses gpt-4-1106-preview, which has a context window of 128k tokens.

\textit{Model Settings.}
To ensure the quality and stability of the LLM's output, we uniformly set the relevant parameters in all experiments of this paper. Among them, the temperature parameter is a hyperparameter that controls the randomness and diversity of text generation. A higher temperature leads to more random and diverse text generation. For consistency, we set the temperature parameter to 0.1.
Top-k and Top-p are two methods for controlling randomness in text generation. Top-k restricts the model to select from the top $k$ candidates with the highest probabilities, ignoring others. In contrast, Top-p selects words whose cumulative probability reaches $p$, for example, when $p = 0.9$, it picks words whose cumulative probability is 90\%. To ensure stability, both Top-k and Top-p are set to their default values, and this setting will apply to all subsequent experiments.

\subsection{Prompting Techniques}
\label{subsec:prompting_techniques}
Recent research~\cite{2025-LLM4CodeSum} on code summarization has demonstrated that even simple zero-shot prompting can instruct LLMs to generate high-quality summaries at the method-level code units. Since this paper primarily focuses on investigating summarization strategies for higher-level code units, we also use zero-shot prompting to instruct the LLMs. This prompting technique is straightforward and less likely to introduce other biases. 

\subsection{Experimental Datasets}
\label{subsec:experimental_datasets}
To ensure the quality of the experiment, we select 10 high-quality Java projects from the GitHub website as the foundation of our experimental dataset. The criteria for selecting these projects are as follows:
1) The proportion of Java language in the project must exceed 90\%. 
2) The project must be ranked within the top 100 Java projects by the number of stars. 
3) The primary language used in the project must be English.
4) We aim to ensure a certain level of diversity among the projects. 

The ten selected projects and their details are shown in Table~\ref{tab:information_of_project_in_experimental_dataset}. 

\begin{table*}[!t] 
    \centering
    \small
    \caption{Information of the ten projects in the experimental dataset.}
    \label{tab:information_of_project_in_experimental_dataset}
    \begin{tabular}{ccl}
        \toprule
        
        Project Name & Star& Description\\
        
        \midrule
        Dubbo
        & 40.3k
        & Dubbo is an open-source, high-performance Java distributed service framework.\\
              
        Apollo & 29.0k & \makecell[l]{Apollo is a reliable configuration management system designed for microservices \\ configuration management scenarios.} \\
         
        Libgdx & 23.0k & libGDX is a cross-platform Java game development framework based on OpenGL (ES).\\
         
        Hadoop & 14.5k
        & Hadoop is an open-source distributed computing framework.\\
        
        Gson & 23.2k & Gson is a Java library that can be used to convert Java objects into their JSON representation.\\
        
        Apktool & 19.5k
        & Apktool is a tool used for reverse engineering third-party, closed, binary Android applications. \\
        
        Elasticsearch & 68.7k
        & \makecell[l]{Elasticsearch is a distributed search and analytics engine optimized for speed and relevance \\ in production-scale workloads.}\\
        
        Nacos & 29.6k
        & \makecell[l]{Nacos is an easy-to-use platform designed for dynamic service discovery, configuration, \\and service management.} \\
        
        Skywalking & 23.5k
        & \makecell[l]{SkyWalking is an open-source APM system that provides monitoring, tracing, \\ and diagnostic capabilities for distributed systems in cloud-native architectures.}\\
        
        Mindustry & 21.7k
        & Mindustry is an open-source tower defense strategy game developed by Anuken.\\
        \bottomrule
    \end{tabular}%
\end{table*}

The ten projects we select are not only widely used and supported by the community but also cover a variety of domains. They include enterprise-level service frameworks (such as Dubbo, Apollo, and Elasticsearch) as well as development tool projects (such as Gson and Apktool). In addition, the selection features the game development framework libGDX and the tower defense game Mindustry.

\section{Results and Findings}
\label{sec:results_and_findings}

\subsection{RQ1: How can high-quality file-level code summaries be effectively generated?}
\label{subsec:answer_to_RQ1}
As mentioned in Section~\ref{subsec:research_questions}, in this RQ, we aim to answer the following two sub-questions through experiments: 1) \textit{Is it necessary to input the full file content into the LLM when generating file-level code summaries?} 
2) \textit{When the number of tokens in a code file exceeds the context window limit of the LLM or exceeds the available budget, what are the alternative strategies for generating file-level code summaries?} 
In the following (sub)sections, we will provide a detailed explanation of the experimental setup and results.

\subsubsection{File Summarization Strategies}
\label{subsubsec:flie_summarization_stragegies}
 
There is relatively little research on summarizing higher-level code units, and few summarization strategies are available for reference. 
To establish a solid research foundation in this field, in this paper, we mainly investigate three simple and easily reproducible code summarization strategies, i.e., full code summarization, reduced code summarization, and hierarchical code summarization. 
As the name suggests, full code summarization refers to inputting all the code involved in the higher-level code unit into the LLMs as context. 
Reduced code summarization refers to inputting only the key information from the code, excluding implementation details such as method bodies. 
Hierarchical code summarization first generates summaries for lower-level code units, and then uses those summaries to generate summaries for higher-level code units. For example, the lower-level code unit for a module-level code unit is the file-level code unit, and the lower-level unit for a file-level code unit is the method-level code unit. 
These three strategies are straightforward and are the most intuitive ones for developers. 
For summarizing different levels of code units, the implementation of these three strategies varies slightly. 
In the following sections, we will specifically introduce the implementation of the three strategies for file-level code unit summarization. 
The use of these three strategies for module-level code unit summarization is discussed in Section~\ref{subsubsec:module_summarization_stragegies}.

\noindent\textbf{1) Full File Code Summarization (FFCS).} 

\begin{figure*}[!t]
    \centering
    \begin{minipage}[t]{0.48\linewidth}
        \centering
        \includegraphics[width=\linewidth]{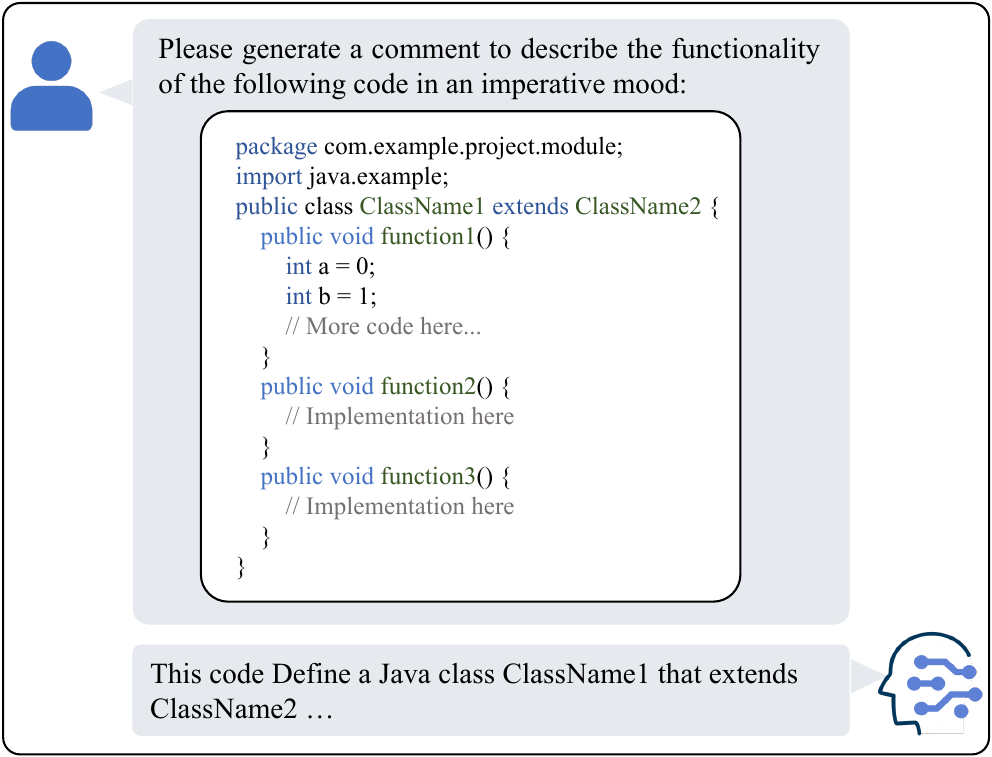}
        \caption{An example of the FFCS strategy.}
        \label{fig:example_of_FFCS}
    \end{minipage}
    \hspace{1mm}
    \begin{minipage}[t]{0.48\linewidth}
        \centering
        \includegraphics[width=\linewidth]{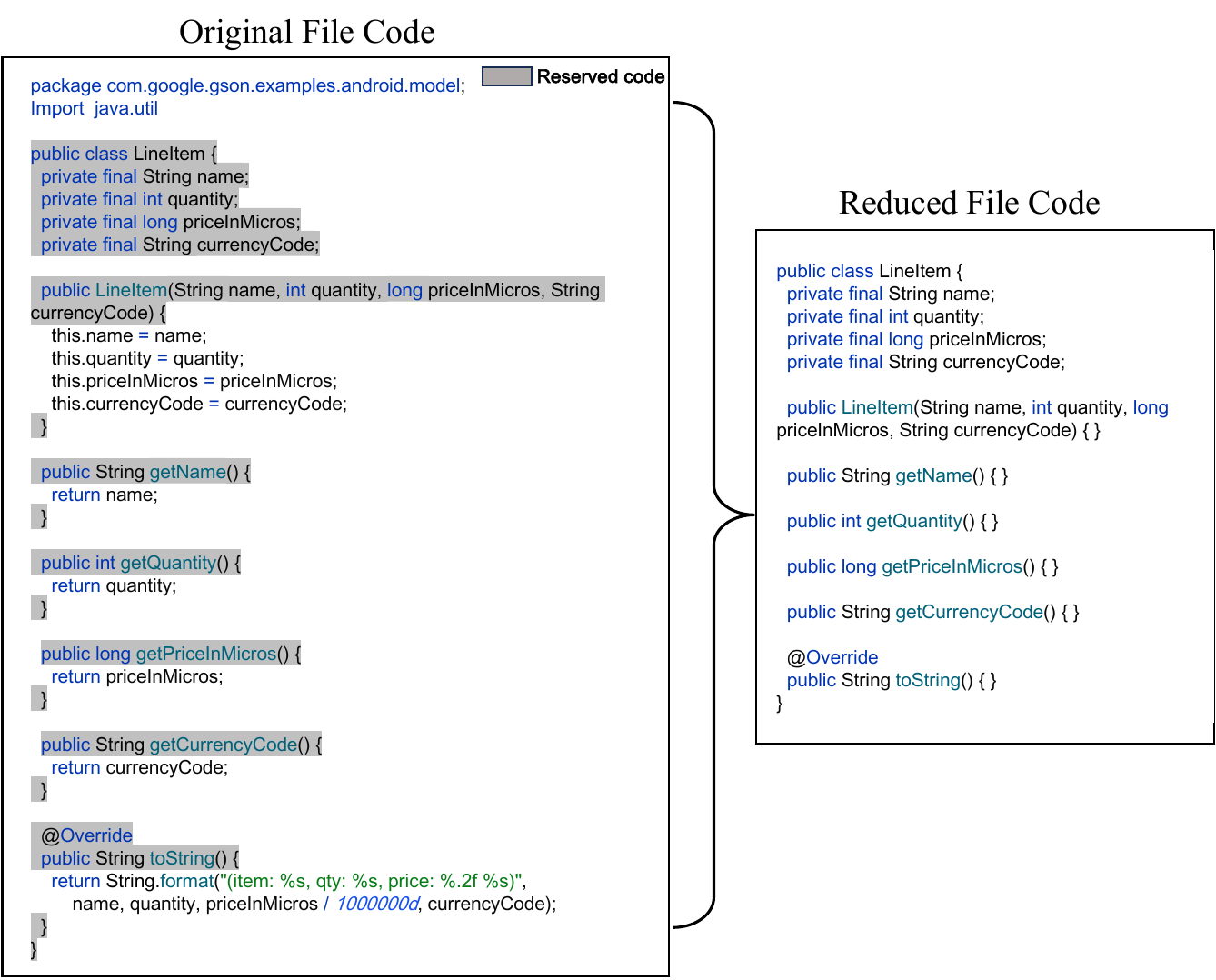}
        \caption{An example of reduced file code.}
        \Description{An example of reduced file code.}
        \label{fig:example_of_reduced_file_code}
    \end{minipage}

\end{figure*}

FFCS is the most direct summarization strategy, but due to the context window limit of LLMs, some longer code units may not be fully entered into the smaller dialogue windows of LLMs. In this strategy, we directly feed the file code to the LLMs to generate summaries. Fig.~\ref{fig:example_of_FFCS} shows the prompt design in the FFCS strategy. 

\noindent\textbf{2) Reduced File Code Summarization (RFCS).}

\begin{figure}[!t]
    \centering
    \includegraphics[width=0.95\linewidth]{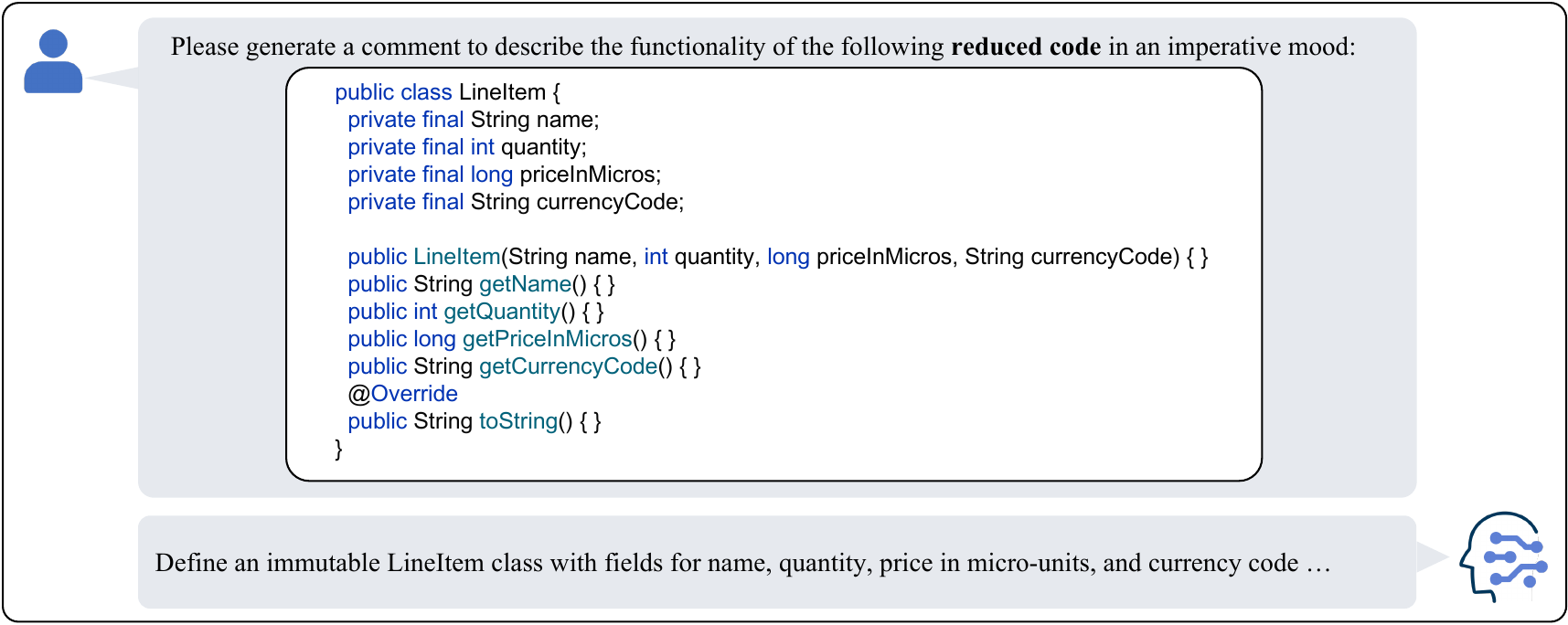}
    \caption{An example of the RFCS strategy.}
    \Description{An example of the RFCS strategy.}
    \label{fig:example_of_RFCS}
\end{figure}

To address the window limits encountered in the FFCS strategy, we refer to relevant studies~\cite{li2023classsum} and propose a reduced file code summarization (RFCS) strategy. 
Fig.~\ref{fig:example_of_reduced_file_code} shows an example of reduced file code. 
In this strategy, we follow~\cite{li2023classsum} and remove the package information, import information, and the implementation details of methods in the file code, i.e., method code bodies. 
Fig.~\ref{fig:example_of_RFCS} shows an example of the RFCS strategy and the prompt design in this strategy.

\noindent\textbf{3) Hierarchical File Code Summarization (HFCS).} 

Another intuitive approach to avoid exceeding the LLM's context window limit with a single input code is hierarchical summarization. For file-level code units, we can first generate summaries for each method-level code unit, and then produce a file code summary based on these method summaries. 

We apply the program slicing technique~\cite{kamkar1995overview, 1984-Program-Slicing} to segment the file code into multiple methods. 
In addition to treating each method as a single method-level code unit, we also explore combining multiple methods into a single method-level code unit. 
We refer to a set of methods combined into a method-level code unit as a method community. 
Therefore, HFCS can be further divided into HFCS\_m and HFCS\_mc, which are summarization strategies for methods and method communities, respectively.

\begin{figure}
    \centering
    \begin{minipage}[t]{0.48\linewidth}
        \centering
        \includegraphics[width=\linewidth]{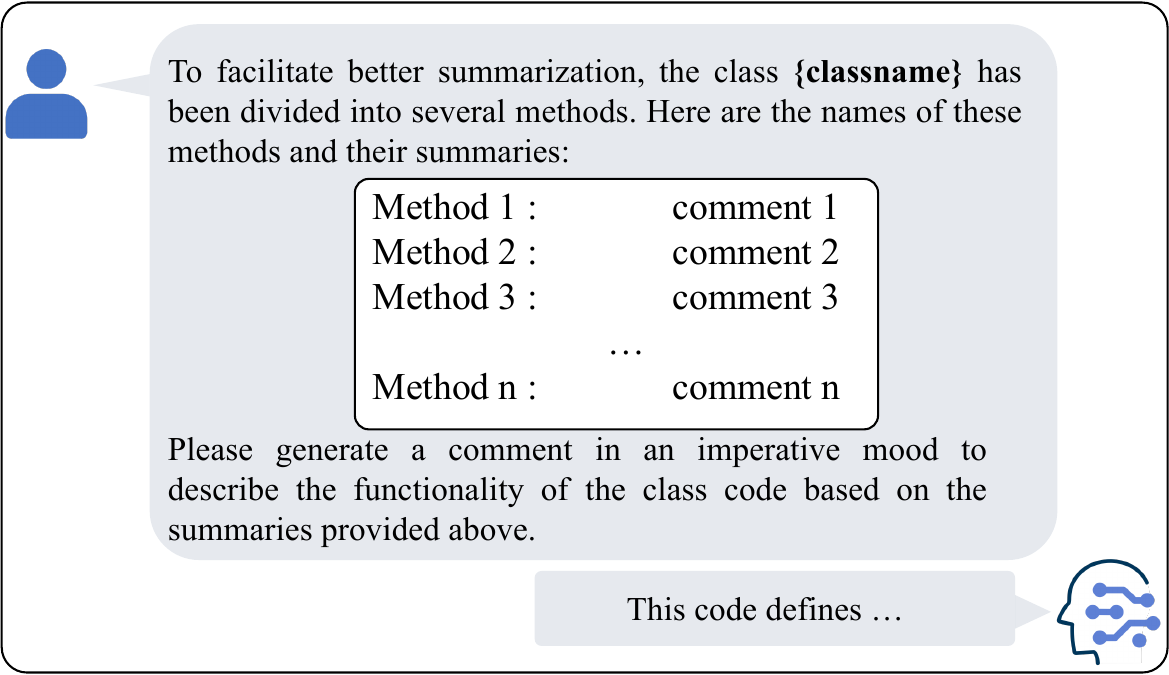}
        \caption{An example of the HFCS\_m strategy.}
        \label{fig:example_of_HFCS_m}
    \end{minipage}
    \hspace{1mm}
    \begin{minipage}[t]{0.48\linewidth}
        \centering
        \includegraphics[width=\linewidth]{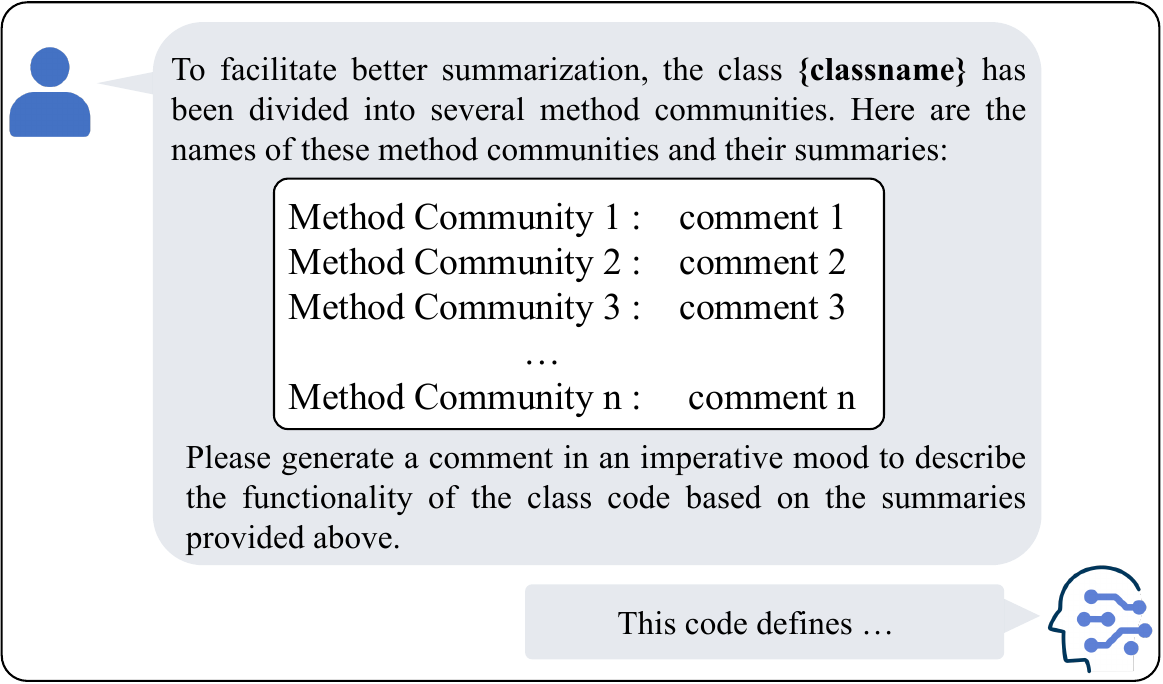}
        \caption{An example of the HFCS\_mc strategy.}
        \Description{An example of the HFCS\_mc strategy.}
        \label{fig:example_of_HFCS_mc}
    \end{minipage}

\end{figure}

\textbf{HFCS\_m.}
In the HFCS\_m strategy, we slice the code of each file according to its methods and retain all global fields used in method calls. Then, we use LLMs to summarize each method. After getting the method summaries, we concatenate all method summaries and use LLMs for further performing summarization, ultimately obtaining a file summary. 
Fig.~\ref{fig:example_of_HFCS_m} shows an example of the HFCS\_m strategy and the corresponding prompt design.

\textbf{HFCS\_mc.}
In the HFCS\_mc strategy, we slice the file code into multiple method communities, each containing multiple methods and global fields. 
In the slicing process, we treat methods and variables as nodes and view the dependencies between methods and variables as edges. Specifically, we use a community detection algorithm~\cite{blondel2008fast} to perform slicing. After slicing, we utilize LLMs to summarize each method community. Based on these communities' summaries, a file summary is further generated. 
Fig.~\ref{fig:example_of_HFCS_mc} shows an example of this strategy.
In our experiments, we find that adding the class name information helps the model predict more accurately. Therefore, we include the class name as the name of the corresponding class in the prompt and then have LLMs further summarize based on this information, ultimately obtaining a file summary.

\subsubsection{Dataset for human evaluation}
\label{sec:dataset-file-level}

The files for evaluation are chosen from 10 famous projects as described in Section~\ref{subsec:experimental_datasets}. To focus on the functionality codes of these projects, test files are excluded by removing any file whose class name or path includes the word ``test''. To ensure a balance between the workload for human evaluators and the representativeness of the results, we randomly select 20 files from each project, resulting in a total of 200 files. The code length of each file does not exceed the context window limits of the three LLMs. For each file, 4 summaries are generated using the four summarization strategies detailed above. This procedure is repeated for each of the three LLMs mentioned in Section~\ref{subsec:experimental_LLMs}, resulting in a total of $3 \times 4 = 12$ summaries per file.

\subsubsection{Human evaluation setup}

To ensure a robust and reliable evaluation of the generated summaries, we invite a group of 5 volunteers, each with over 3 years of professional experience in software development and strong proficiency in English. This ensures that the evaluators have the necessary expertise to assess the quality of the summaries in terms of both technical accuracy and language fluency.

For the evaluation process, a total of 200 selected files are used, with each file paired with 12 generated summaries. To mitigate potential biases, the summaries for each file are randomly shuffled before presentation to the evaluators. Volunteers are provided with the source code of each file along with the corresponding summaries, ensuring that they have a clear understanding of the code context before evaluating the summaries.

Following the evaluation framework proposed by Shi et al.~\cite{2022-Evaluation-Neural-Code-Summarization}, each volunteer is asked to rate the summaries on a scale from 1 to 5, where a higher score corresponds to better summary quality. The scoring criteria are as follows:
\begin{itemize} 
\item \textbf{1 point}: The summary is severely flawed, such as being completely inconsistent with the code content or omitting essential functionality and architectural details.
\item \textbf{2 points}: The summary contains significant flaws, such as covering only part of the content and lacking clarity.
\item \textbf{3 points}: The summary is generally accurate but lacks completeness or clarity, reflecting only part of the code's key aspects.
\item \textbf{4 points}: The summary is high quality, mostly accurate, and complete, with only minor areas for improvement.
\item \textbf{5 points}: The summary is nearly flawless, comprehensively and clearly reflecting the code's core functionality, architecture, and design intent, with strong readability and expression.
\end{itemize}

This evaluation scale is designed to cover a range of summary qualities, allowing the evaluators to assess the summaries with precision and consistency. To further improve the objectivity of the evaluation, multiple evaluators are involved in assessing each summary, which helps mitigate any potential biases introduced by individual assessors.

Additionally, as an incentive, volunteers who complete the evaluation task are awarded course credits. Volunteers are also provided with background information about the selected projects, including the project name, core functionality, and the specific modules or file code being evaluated. This context enables the evaluators to better understand the purpose of the code and assess the quality of the generated summaries more accurately.

\subsubsection{Evaluation Results}

\begin{table}[!t]
    \small
    \caption{Performance of LLMs with the four summarization strategies on commenting file-level code units.}
    \tabcolsep=8pt
    \label{table:file-avg-score}
    \begin{tabular}{lllllll}
        \toprule
        & \multirow{2}{*}{FFCS} & \multirow{2}{*}{RFCS} & \multicolumn{2}{c}{HFCS} & \multirow{2}{*}{Average} \\

        \cmidrule(lr){4-5}
        
        & & & HFCS\_m & HFCS\_mc  & \\ 
        
        \midrule
        
        GPT-4      & 4.737 ($\pm$0.016) & 3.625 ($\pm$0.035) & 3.396 ($\pm$0.031) & 3.131 ($\pm$0.031) & 3.722 \\
        
        CodeLlama & 4.468 ($\pm$0.023) & 3.297 ($\pm$0.030) & 3.445 ($\pm$0.028) & 3.242 ($\pm$0.031) & 3.613 \\
        
        CodeGemma & 4.188 ($\pm$0.028) & 3.276 ($\pm$0.033) & 3.150 ($\pm$0.034) & 3.132 ($\pm$0.034) & 3.436 \\ 
        
        \midrule
        
        Average & 4.464 & 3.399 & 3.330 & 3.168              & /     \\ 
        
        \bottomrule
    \end{tabular}
\end{table}

The average scores by human evaluation for the summaries generated by the three LLMs with the four summarization strategies are presented in Table~\ref{table:file-avg-score}. It is observed that all three LLMs produced better summaries when provided with the full code compared to summaries generated from code compressed by any of the three strategies. On average, the score for full code summaries is 31.33\% higher than the best score achieved using the compression strategies. It means that compression adversely affects the LLMs' ability to interpret the file code.

To validate the statistical significance of this finding, we conduct one-sided Wilcoxon-Mann-Whitney tests~\cite{arcuri2014hitchhiker, 2015-Practical-Test-Selection} to determine whether summaries generated based on the full code receive significantly higher scores than those generated based on the compressed code. The p-value ($p<10^{-30}$) is extremely small, indicating that the difference is statistically significant, as a p-value below 0.05 typically suggests strong evidence against the null hypothesis. This confirms that summaries generated from the full code perform significantly better than those from the compressed code.

\finding{LLMs generate better summaries for code files when provided with the full code compared to compressed code inputs.}

\begin{figure*}[!t]
    \centering
    \includegraphics[width=0.7\linewidth]{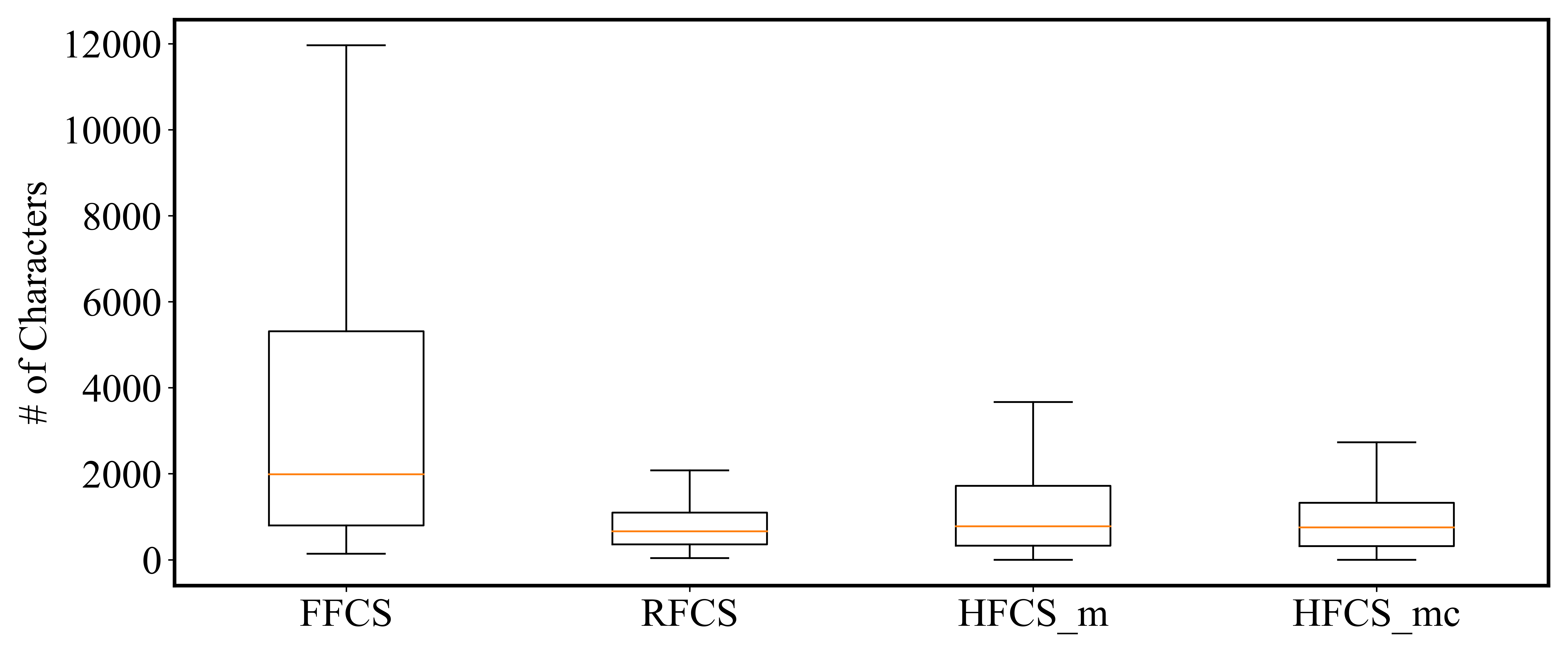}
    \caption{Distribution of the number of characters in the input produced by different summarization strategies.}
    \Description{Distribution of the number of characters in the input produced by different summarization strategies.}
    \label{fig:file-size-all-methods}
\end{figure*}

Although using the full code for file-level summarization yields the best results, it also introduces significantly higher costs. 
Fig.~\ref{fig:file-size-all-methods} shows the distribution of file sizes for the full code and those compressed by various strategies, where size is measured by the number of characters since the three LLMs use different tokenizers. As shown in the figure, the full code is 294.4\% longer than the code compressed by \declcomp, which produces the shortest compressed code. The \funcslice and \commslice strategies also reduce the size of the code to 39.40\% and 30.21\%, respectively.

When comparing the three compression strategies, it can be found that \commslice generally produces the worst results: the quality of the generated summaries is the lowest according to human evaluation, and it is also the most costly one among the three compression strategies.
For \declcomp and \funcslice, \declcomp generally performs better because it 1) outperforms \funcslice in summary quality in two out of three LLMs, 2) produces shorter compressed code, and 3) does not require first summarizing the slices. However, it is also worth noting that this is not the case for CodeLlama. For CodeLlama, the \funcslice strategy yields better results than \declcomp, with a 4.49\% higher score.

To ensure the robustness of our findings, we conduct Wilcoxon-Mann-Whitney tests on the evaluation scores of the summaries produced by the different strategies for GPT4, CodeGemma, and CodeLlama. The results show that for GPT4 and CodeGemma, \declcomp significantly outperforms \funcslice and \commslice in producing higher-quality summaries at a lower cost ($p_1 = 4.779\times 10^{-8}$, $p_2 = 4.060 \times 10^{-26}$, $p_3 = 0.003$, $p_4 = 0.001$). On the other hand, for CodeLlama, \funcslice generates better summaries than the other strategies ($p_1 = 0.001$, $p_2 = 1.05 \times 10^{-5}$). These results confirm the statistical significance of the observed differences, supporting our conclusions that \declcomp is more effective for GPT4 and CodeGemma, while \funcslice is superior for CodeLlama. The significant findings presented thereafter will undergo Wilcoxon-Mann-Whitney, and the corresponding data will be explicitly reported in the paper.

\finding{For GPT4 and CodeGemma, \declcomp outperforms \funcslice and \commslice by producing higher quality summaries at a lower cost. For CodeLlama, \funcslice produces better summaries.}

When comparing the three LLMs, GPT4 generally outperforms the two code LLMs, CodeLlama ($p<10^{-30}$) and CodeGemma ($p<10^{-30}$), when summarizing using full code or \declcomp. Specifically, GPT4's average scores are 6.02\% and 9.95\% higher than those of CodeLlama for full code and \declcomp summaries, respectively. Meanwhile, CodeLlama outperforms CodeGemma by 6.69\% and 0.64\% on average for these two strategies.
However, for the two hierarchical summary approaches, GPT4 does not maintain its lead. In these cases, CodeLlama leads GPT-4 ($p_1=0.178$, $p_2=0.001$) and CodeGemma ($p_1<10^{-30}$, $p_2=0.009$) in HFCS\_m and HFCS\_mc. Although the statistical test in HFCS\_m does not show a particularly significant advantage over GPT4, CodeLlama still has the upper hand in terms of mean scores. Overall, CodeLlama performs the best in the two hierarchical summary approaches.

\finding{For full code and \declcomp approaches, GPT-4 significantly outperforms CodeLlama and CodeGemma. For the two hierarchical approaches, CodeLlama leads the other models.}

\summary{Combining the previous results, we find that when both the context window and budget are sufficient, the optimal strategy is to input the full code into the LLM, as it yields the highest quality summary. However, if budget constraints exist, using \declcomp with a commercial model like GPT-4 can reduce token usage by approximately 74.65\% with only a 23.47\% reduction in summary quality. The two \hier strategies are not recommended for file-level summarization since they use more tokens than the other two strategies but produce worse results.}

\subsection{RQ2: How can high-quality module-level code summaries be effectively generated?}

Similarly, as mentioned in Section~\ref{subsec:research_questions}, in this RQ, we aim to answer the following two sub-questions through experiments: 1) \textit{Is it necessary to input the entire module content into the LLM when generating module-level code summaries?} 
2) \textit{When the number of tokens in a module exceeds the context window limit of the LLM or exceeds the available budget, what are the alternative strategies for generating module-level code summaries?}  
In the following (sub)sections, we will provide a detailed explanation of the experimental setup and results.

\subsubsection{Module Summarization Strategies.} 
\label{subsubsec:module_summarization_stragegies}
As mentioned in Section~\ref{subsubsec:flie_summarization_stragegies}, we also compare the three summarization strategies for module-level code summarization.

\noindent\textbf{1) Full Module Code Summarization (FMCS).} 
Similar to the file-level summarization, we also consider inputting the full module code to LLMs, if the code does not exceed the context window limit. For clarity, we refer to this module-level summarization strategy as Full Module Code Summarization (FMCS).

\noindent\textbf{2) Reduced Module Code Summarization (RMCS).} 
RMCS is similar to FMCS and removes all method code bodies from the module code. Unlike RFCS, RMCS may include multiple reduced file code units.

\noindent\textbf{3) Hierarchical Module Code Summarization (HMCS).}

HMCS adopts the same hierarchical approach as HFCS. The difference is that HMCS only involves two levels, namely the file level and the module level, without involving the method level. 
In short, HMCS first generates summaries for each file code, and then produces a module summary based on these file code summaries.

\subsubsection{Dataset for human evaluation}

In this paper, we identify the modules based on the Java packages. Then for each of the 10 projects, as detailed in Table~\ref{tab:information_of_project_in_experimental_dataset}, we randomly select 5 modules.  Therefore, a total of 50 modules are selected. Then for each of the modules, we generate summaries using the three strategies described above and the three LLMs. So a total of $9=3\times3$ summaries will be generated for each of the modules.

\subsubsection{Human evaluation setup}
\label{subsubsection:human_evaluation_setup_in_module_summarization}

To evaluate the generated summaries, we also invite 5 volunteers, each with over 3 years of software development experience and excellent English ability. Among the 50 modules, considering that manually evaluating all of them would be challenging due to the large number, we select the modules with at most 10 files for human evaluation, resulting in 17 modules. Each module contains 9 summaries for evaluation. However, not all the summaries can be generated due to the limited context window of LLMs. If a summary fails to be generated, we will leave it blank and ask the volunteers not to rate them. Then, we provide the files of each module along with the corresponding summaries and ask volunteers to rate the quality of the summaries on a scale from 1 to 5, with higher scores indicating better quality. The specific rating criteria are similar to those used for the file-level evaluation.

\subsubsection{Evaluation Results}

\begin{table}[!t]
    \centering
    \scriptsize
    \begin{minipage}[c]{0.48\textwidth}
        \tabcolsep=6pt
        \caption{Performance of LLMs with three summarization strategies on commenting module-level code units. The values in parentheses represent the standard deviation. AVG: Average.}
        \label{table:module-avg-score-1}
        \begin{tabular}{lcccc}
            \toprule
            & FMCS & RMCS & HMCS & AVG \\ 
            
            \midrule
            
            GPT-4 & 2.706 ($\pm$0.081) & 2.647 ($\pm$0.081) & 3.071 ($\pm$0.079) & 2.808 \\
            
            CodeLlama & 3.000 ($\pm$0.165) & 3.293 ($\pm$0.106) & 3.600 ($\pm$0.091) & 3.298 \\
            
            CodeGemma & 2.565 ($\pm$0.155) & 3.671 ($\pm$0.080) & 3.871 ($\pm$0.074) & 3.369 \\ 
            
            \midrule
            
            Average & 2.757              & 3.204              & 3.514 &       \\ 
            \bottomrule
        \end{tabular}
    \end{minipage}
    \hfill
    \begin{minipage}[c]{0.48\textwidth}
        \tabcolsep=6pt
        \caption{Performance of LLMs with three summarization strategies on commenting module-level code units (removing meaningless summaries). The values in parentheses represent the standard deviation.}
        \vspace{-4mm}
        \label{table:module-avg-score-2}
        \begin{tabular}{lcccc}
            \toprule
            
            & FMCS & RMCS & HMCS & AVG \\ 
            
            \midrule
            
            GPT-4      & 2.640 ($\pm$0.097) & 2.600 ($\pm$0.094) & 2.980 ($\pm$0.100) & 2.740 \\
            
            CodeLlama & 3.314 ($\pm$0.170) & 3.333 ($\pm$0.126) & 3.600 ($\pm$0.111) & 3.415 \\
            
            CodeGemma & 3.660 ($\pm$0.104) & 3.600 ($\pm$0.098) & 3.940 ($\pm$0.096) & 3.733 \\ 
            
            \midrule
            
            AVG       & 3.205              & 3.178              & 3.506              &       \\ 
            
            \bottomrule
        \end{tabular}
    \end{minipage}
\end{table}

Since the context window of LLMs is limited, not all summaries can be generated for all 17 modules.
Among the 17 modules, 5 exceed the window size of CodeGemma, and 3 exceed the window size of CodeLlama. 
Therefore, a total of $17 \times 9-7-2 = 145$ summaries are successfully generated for the 17 modules. 

The averaged human evaluation score along with the standard errors are shown in Table~\ref{table:module-avg-score-1}. 
It can be found that the FMCS strategy, which achieves the best performance in summarizing files, performs the worst in summarizing modules. 
It has 13.98\% and 21.54\% lower scores when compared to the RMCS and HMCS strategies. 

\begin{figure}[!t]
    \centering
    \includegraphics[width=0.95\linewidth]{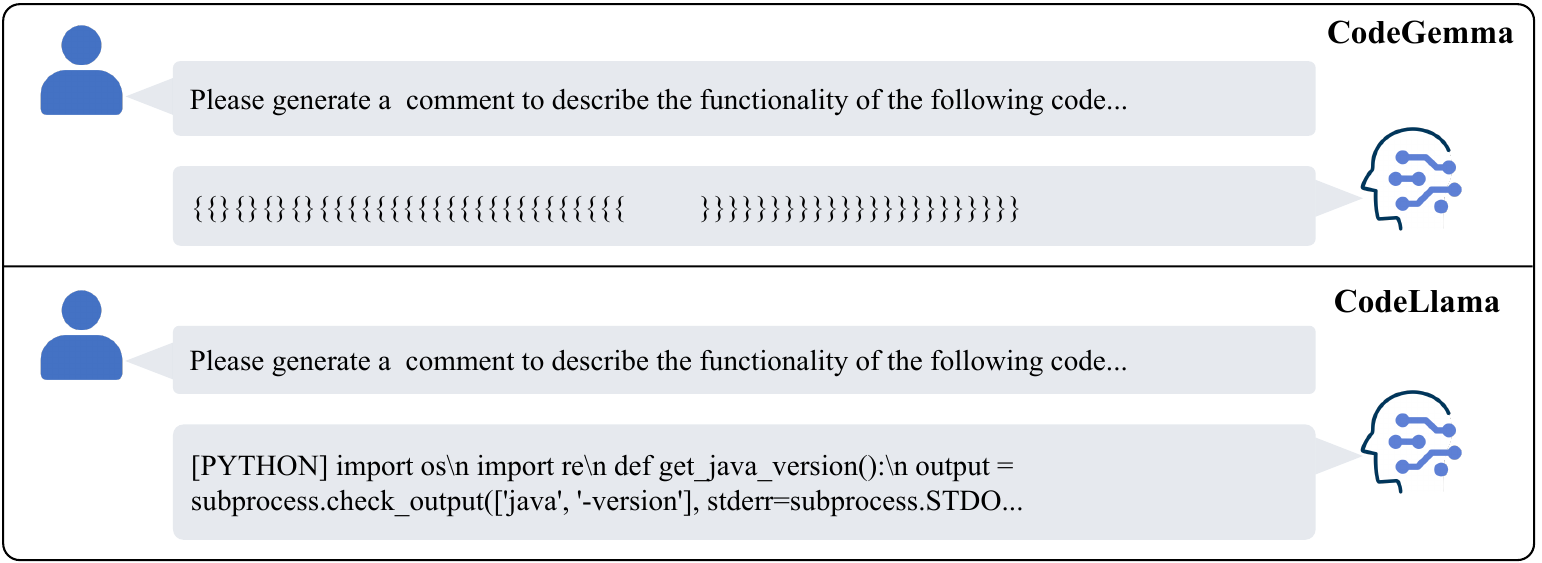}
    \caption{Examples of meaningless summaries generated by CodeGemma and CodeLlama with FMCS.}
    \Description{Examples of meaningless summaries generated by CodeGemma and CodeLlama with FMCS.}
    \label{fig:llmerr}
\end{figure}

We further investigate the reason for the FMCS strategy's low score. 
We find that when the prompt is long, CodeLlama and CodeGemma often generate meaningless summaries, even though the prompt is shorter than their window size. The meaningless summaries generated by each of the models have their own pattern. 
We show two typical examples of meaningless summaries in Fig.~\ref{fig:llmerr}. For CodeGemma, it sometimes generates summaries that repeat some special characters like spaces or '\}'. For CodeLlama, it generates some Python codes that have no obvious relation to the input module code.

We manually examined all 145 summaries and found that only FMCS strategy may generate meaningless summaries, which could be due to its long context. For 2 modules, both CodeLlama and CodeGemma generated meaningless summaries. And for another 5 modules, only CodeGemma generated meaningless summaries. 

\finding{CodeGemma and CodeLlama sometimes fail to generate meaningful summaries when using the FMCS strategy.}

To compare the quality of non-meaningless summaries, we exclude the 7 modules where some LLMs generated meaningless summaries. Then the averaged score for the remaining 10 modules is shown in Table~\ref{table:module-avg-score-2}. 

It can be found that except for the scores for CodeLlama and CodeGemma when using the FMCS strategy (i.e., the scenario that LLMs generate some meaningless summaries), the average human evaluation score roughly remains stable. 

Based on the updated scores, we can find that even after excluding the meaningless cases, the FMCS strategy still generates summaries that are worse than the RMCS and HMCS strategies. On average, FMCS's score is 0.84\% and 9.36\% lower than the RMCS and HMCS strategies, respectively. This result is in contrast to the case for file summarization where using FMCS can produce much better summaries than the RMCS and HMCS strategies. 

To see the possible reasons, for each module, we calculated its averaged FMCS score. Then we find that the score decreases fastly as the module size increases. We then measured the correlation between the averaged score and the length of the module, the result shows that they have a correlation of -0.847, which means they are strongly negatively related.
Such a correlation indicates that as the modules grow larger, FMCS may produce worse results.

In contrast to the poor performance of FMCS, we can find that HMCS has the best performance in module summarization. It has 9.39\% and 10.32\% higher scores on average than RMCS and FMCS strategies, respectively. 

In the statistical tests, the HMCS strategy significantly outperforms FMCS ($p=0.010$) and RMCS ($p=0.002$) in GPT-4 evaluations, and also outperforms FMCS ($p=0.031$) and RMCS ($p=0.008$) in CodeGemma evaluations. For CodeLlama, although the results are not statistically significant when comparing HMCS with FMCS ($p=0.188$) and RMCS ($p=0.061$), HMCS still shows an advantage in terms of mean scores.

\finding{After excluding meaningless summaries, FMCS is not significantly different from RMCS. Meanwhile, The HMCS strategy performs significantly better than FMCS and RMCS.}

\begin{figure}[!t]
    \centering
    \begin{minipage}[c]{0.25\linewidth}
        \includegraphics[width=\linewidth]{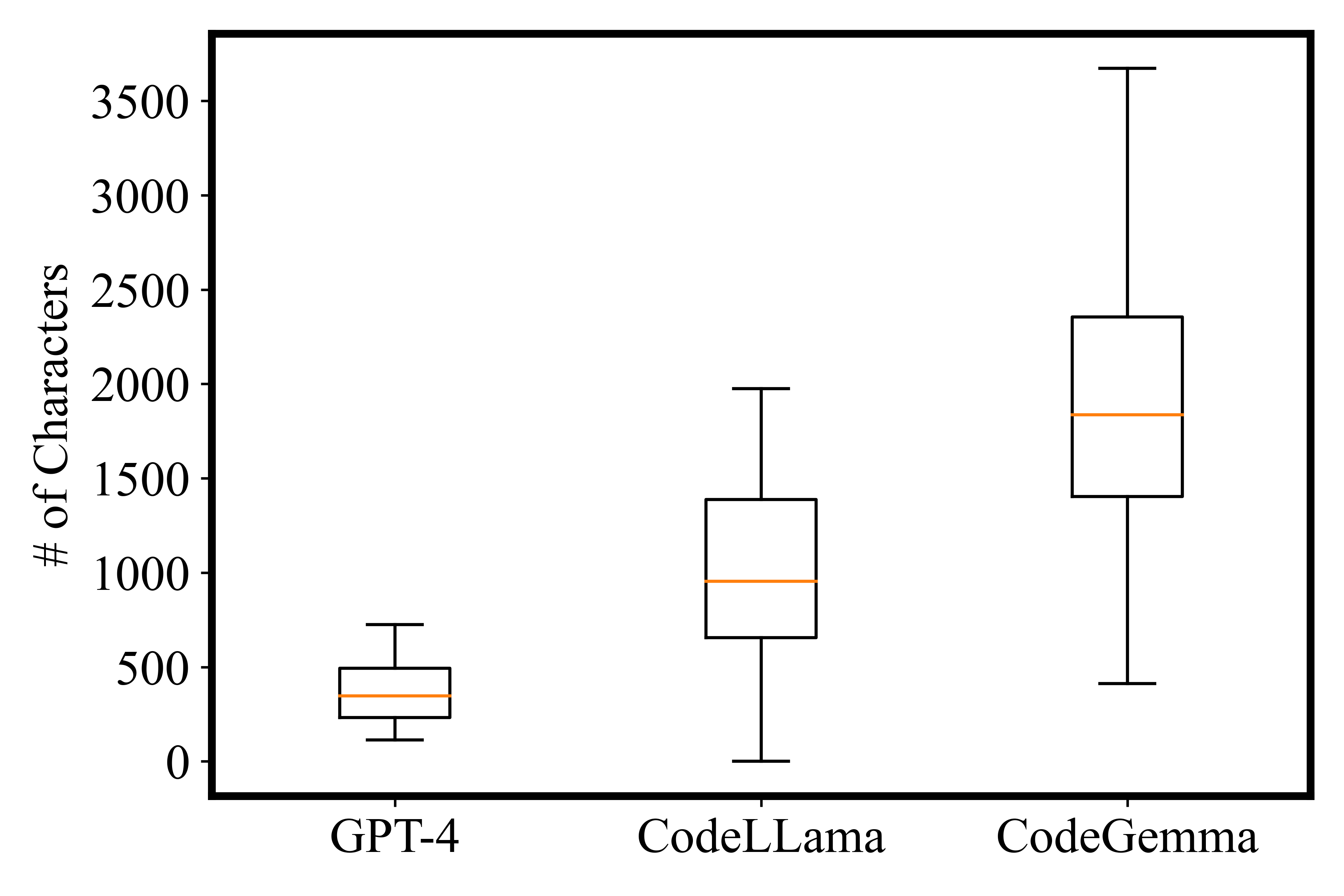}
        \caption{Distribution of characters in module summaries generated by the three LLMs.}
        \label{fig:module-comment-len}
    \end{minipage}
    \hfill
    \begin{minipage}[c]{0.72\linewidth}
        \centering
        \includegraphics[width=\linewidth]{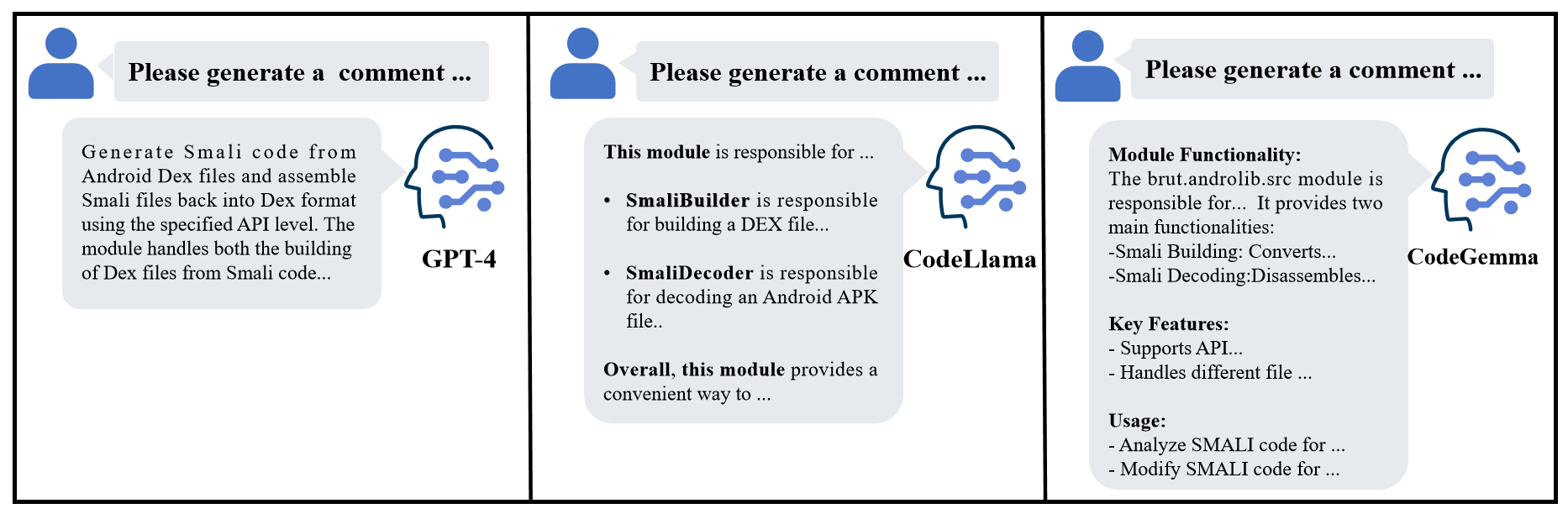}
        \caption{Examples of module summaries generated by the three LLMs.}
        \Description{Examples of module summaries generated by the three LLMs.}
        \label{fig:llm-module-result-case}
    \end{minipage}
\end{figure}

Meanwhile, it is worth noting that CodeGemma has a higher score than CodeLlama ($p = 0.002$) and GPT-4 ($p = 3.514 \times 10^{-24}$), while GPT-4 has the lowest score.

This contradicts our conclusion on file-level summary that GPT-4 has the best performance. 
We further investigate the reason and find that it could be due to the reason that GPT-4 is too lazy. 
We show the distribution of module-level summaries in Fig.~\ref{fig:module-comment-len}. It can be found that the module summary generated by GPT-4 is much shorter than that generated by CodeGemma and CodeLlama by having a median value of only 346 characters, while this number for CodeGemma and CodeLlama is 995 and 1,836, respectively. To further understand their difference, we manually inspected the generated summaries. We find that each of the LLM has its own pattern in generating summaries. 
We show three typical cases in Fig.~\ref{fig:llm-module-result-case}. For GPT-4, it commonly generates only one short paragraph to summarize the module. For CodeGemma, it often summarizes the modules with several sections, each focusing on a specific aspect like functionalities or usages. For CodeLlama, it often first summarizes the overall modules, then introduces sub-classes, and finally concludes its summarization. Both of the formats of CodeLlama and CodeGemma can provide a more detailed and clear understanding to the module code than GPT4's single-paragraph answer. This could be the reason that GPT-4 has a lower score in module-level summarization.

\finding{For module-level summarization, GPT-4 generates much shorter summaries than CodeLlama and CodeGemma. Based on the human evaluation, CodeGemma is the best model for module-level summarization.}

\summary{Combining the result in this RQ, we can find that the \hier is the best strategy for module-level summarization by producing better summaries with shorter context. However, considering that \hier needs to first summarize all files, it would cost a lot of tokens. To lower the cost, the best alternative should be the RMCS strategy as it produces slightly better summaries than FMCS with a much lower cost.\\When choosing LLM for module-level summarization, it would be better to use CodeGemma when the module is small since it can produce better summaries. However, considering its limited window size, for larger modules, GPT-4 could be the only choice since it has a much larger context window.}

Note that in this RQ, our findings are based on the 17 modules that have no more than 10 files since it is hard for human evaluators to evaluate large modules that contain too many files. In the next RQ, we will investigate if the LLMs are also qualified as evaluators, and see if they can be used to evaluate the summaries generated for larger modules.

\subsection{RQ3: How can the quality of file-level and module-level code summaries be automatically evaluated?}

In this RQ, we aim to 1) investigate the feasibility of using LLMs as evaluators, and 2) if feasible, is it possible to use LLMs to evaluate the large modules that are hard to evaluate by humans. Since we want to use LLMs to evaluate large modules, we only evaluate if GPT-4 is qualified as an evaluator since CodeLlama and CodeGemma have limited context window sizes. 
To evaluate if GPT-4 is qualified as an evaluator, we let it evaluate summaries for both file and module summaries to see if its result agrees with humans.

\subsubsection{GPT-4 in Evaluating File-level Summary}
First, we let GPT-4 evaluate the summaries that are evaluated by humans. Therefore, the data selected is identical to that we detailed in Section~\ref{sec:dataset-file-level}.

\noindent\textbf{Experimental Setup.}

\begin{figure*}[!t]
    \centering
    \includegraphics[width=0.95\linewidth]{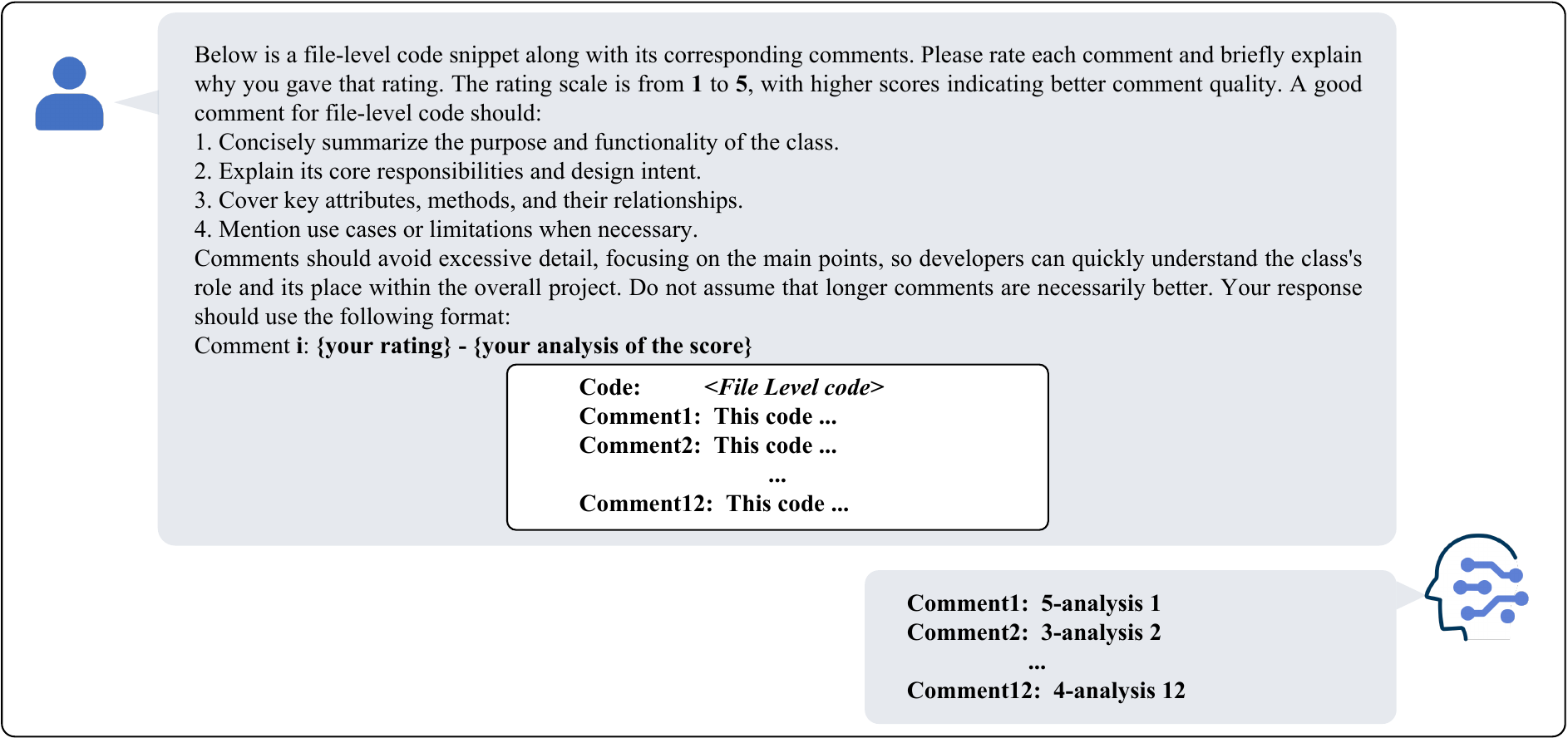}
    \caption{Example of evaluating file summaries using GPT-4.}
    \Description{Example of evaluating file summaries using GPT-4.}
    \label{fig:fileevl}
\end{figure*}

The steps for LLM evaluation are similar to those of human evaluation. Specifically, for each sample, we provide detailed scoring criteria and ask the LLM to evaluate the sample on a scale of 1 to 5, accompanied by corresponding analysis. Higher scores represent higher quality. The prompt used for scoring by the large model is shown in Fig.~\ref{fig:fileevl}.

We use the LLM to score 200 file-level cases and provide the module information using the three different file-level summarization strategies. In total, we conduct $200 \times 4=800$ evaluations.

\noindent\textbf{Measuring the Agreement.}
We follow~\cite{2022-Evaluation-Neural-Code-Summarization, 2025-LLM4CodeSum} to use Spearman's correlation coefficient to measure the Agreement between humans and GPT's scores. Spearman's correlation is a measure that assesses the strength and direction of a relationship between two ranked variables. Its values range from -1 to +1, where -1 indicates a perfect negative correlation, +1 indicates a perfect positive correlation, and 0 signifies no correlation.

\begin{table*}[!t]
    \centering
    \footnotesize
    \tabcolsep=2.5pt
    \caption{Comparison of scores by human and GPT-4 for the summaries of file-level code units.}
    \label{table:file-corr-file}
    \begin{tabular}{lccccccccccccc}
    
    \toprule
    
    & \multicolumn{3}{c}{\multirow{2}{*}{FFCS}} & \multicolumn{3}{c}{\multirow{2}{*}{RFCS}} & \multicolumn{6}{c}{HFCS} & \multirow{3}{*}{Correlation} \\ 

    \cmidrule(lr){8-13} 
    
    & & & & & & & \multicolumn{3}{c}{HFCS\_m} & \multicolumn{3}{c}{HFCS\_mc} & \\ 
    
    \cmidrule(lr){2-4} \cmidrule(lr){5-7} \cmidrule(lr){8-10} \cmidrule(lr){11-13}
    
    & GPT-4 & Human & Correlation & GPT-4 & Human & Correlation & GPT-4 & Human & Correlation & GPT-4 & Human & Correlation & \\
    
    \midrule
    
    GPT-4 & 3.872& 4.737& 0.154& 2.895& 3.625& 0.740& 2.735& 3.396& 0.573 &  2.500& 3.131& 0.535&  0.724 \\
    
    CodeLlama & 3.760& 4.468& 0.271& 2.316& 3.297& 0.578& 2.732& 3.445& 0.433& 2.439& 3.242& 0.495& 0.623 \\
    
    CodeGemma & 3.768& 4.188& 0.197& 2.770& 3.276& 0.555 & 2.427& 3.150& 0.420& 2.205& 3.132& 0.434 &  0.620 \\
    
    \midrule
    
    Correlation & \multicolumn{3}{c}{0.185 }& \multicolumn{3}{c}{0.626 }& \multicolumn{3}{c}{0.491 }& \multicolumn{3}{c}{0.483 }& \\
    
    \bottomrule
    \end{tabular}
\end{table*}

\noindent\textbf{Results.}
To see if the GPT-4 result agrees with the human evaluation, we list the average score from both human and GPT-4 in Table~\ref{table:file-corr-file}. 
By comparing the scores, we can find that the average score from GPT-4 is lower than that from humans ($p < 10 \times 10^{-30}$). The average score from GPT-4 is 2.868, while the number is 3.591 for humans. Though they have differences in the average score, the trend of scores is very similar to that of humans, which means the qualitative conclusions drawn from GPT-4 and human scores would be the same. For example, both GPT-4 and humans believe that the summary generated by the FFCS strategy is better than that by the RFCS strategy or the two HFCS strategies. 

To quantitatively measure the agreement, for each type of summary (identified by the combination of the LLM and summarization strategy), we measure the correlation and between GPT's score and human's score. Moreover, we measure the overall correlation on each LLM or summarization strategy, which is shown in the last line and column of Table~\ref{table:file-corr-file}.
Based on the result, we can find that in most cases, the GPT's overall score is correlated to humans by having a correlation between 0.420 and 0.740. The only exception is the scores for the FFCS strategy. It has an overall correlation of only 0.185. We manually investigate the cases and find the main reason is that humans tend to give 5 (i.e. the highest score) for a comment if it is overall sound by precisely describing the code's functionality. However, GPT-4 is much harsher: it refuses to give a full score if the comment has any subtle imperfection like the comment is overly long by having three sentences. 

\finding{For file-level summary, the average score from GPT-4 is lower than that from humans. 

Regardless of whether scores come from human evaluators or GPT-4, the FFCS strategy performs better than RFCS and both HFCS strategies.
The GPT-4 score has a moderate or strong correlation to the human score in most cases. The exception is that when the comment's quality is high, human tends to give it the highest score, while GPT-4 is harsher on summaries' minor imperfections}

\subsubsection{LLM in Evaluating Module-level Summary}
For file level, we find that the qualitative comparison results concluded from the GPT-4's scores are the same as that concluded from the human scores. In this section, we aim to investigate if the situation is the same for module-level summaries.

\noindent\textbf{Experimental Setup.}

The experiment setup is overall the same as the file-level experiment. The only difference is that some modules exceed the limitations of LLMs, we leave the score empty when the summary is not available. We provide module information using the three module summarization strategies, conducting a total of $50 \times 3 = 150$ evaluations.

When evaluating the module-level summaries with GPT-4, there are several key differences from the setup of the file-level summaries. First, for file-level summaries, we provide GPT-4 with the full code for it to evaluate the summaries. However, for the module level, the full code could be too long and it could fail to fit in GPT-4's context window size. Moreover, according to our result in RQ2, providing full module code may not be beneficial for GPT-4 to understand the module. Therefore, in this experiment, we also evaluate if GPT-4 can evaluate the summaries based on 1) full module code (i.e., FMCS), 2) reduced module code (i.e., RMCS), or 3) hierarchical module code summaries (i.e., HMCS). In other words, we deploy 3 GPT agents: GPT-F, GPT-R, and GPT-S, where GPT-F takes the full module code to evaluate the summaries, GPT-R takes the reduced module code, and GPT-S takes hierarchical module code summaries (i.e., the file summaries). Since we have 3 GPT agents to evaluate 50 modules each with 9 summaries, a total of $3 \times 50 \times 9 = 1350$ scores are marked by GPT-4.

\noindent\textbf{Comparison to Human Evaluation.}
First, we compare the GPT-4's evaluation score to the scores marked by human evaluators on the 17 modules that are also evaluated by humans in RQ2. 
The results are reported in Table~\ref{table:module-gpt-vs-hm}. Based on the raw scores, we can find that the GPT agents give a much higher score to the GPT-4-generated summaries, while such bias does not exist for the summaries generated by CodeLlama and CodeGemma. This indicates that for module-level summaries, the GPT agents prefer the GPT-4-generated summaries ($p = 6.516 \times 10^{-7}$). Therefore, unlike the situation of file level, it might not be fair to compare the performance of different LLMs based on the GPT agents' scores. 

However, from the table, we can find that the correlation on the right side, i.e., the correlation between human and GPT evaluator on the result from the same LLM, is still relatively high by ranging from 0.302 to 0.882. This correlation suggests that GPT evaluators are capable of assessing summaries produced by the same LLMs, despite differing summary generation strategies.

\finding{Compared with humans, GPT agents prefer the summaries generated by GPT-4. However, when comparing summaries generated by the same LLM, GPT agents have a high correlation with humans.}

\begin{table}[!t]
    \centering
    \small
    \caption{Comparison of scores by human and GPT-4 for the summaries of module-level code units.}
    \label{table:module-gpt-vs-hm}
    \resizebox{\textwidth}{!}{ 
        \begin{tabular}{lccccccccccccccc}
        
            \toprule
            
            & \multicolumn{4}{c}{FMCS} & \multicolumn{4}{c}{RMCS} & \multicolumn{4}{c}{HMCS}& \multicolumn{3}{c}{Correlation} \\ 
            
            \cmidrule(lr){2-5} \cmidrule(lr){6-9} \cmidrule(lr){10-13} \cmidrule(lr){14-16}
            
            & GPT-F & GPT-R & GPT-S & Human & GPT-F & GPT-R & GPT-S & Human & GPT-F & GPT-R & GPT-S & Human & GPT-F & GPT-R & GPT-S  \\
            
            \midrule
            
            GPT-4 & 4.250& 4.353& 3.882& 2.706& 4.000& 4.059& 3.588& 2.647& 4.438& 4.353& 4.588& 3.071&  0.302& 0.364&0.659 \\
            
            CodeLlama & 2.000&  2.214& 2.154& 3.000& 2.867& 3.125& 2.688& 3.293& 4.188& 4.294& 4.812& 3.600& 0.538& 0.578&0.573 \\
            
            CodeGemma & 2.467& 2.312& 2.375& 2.565& 3.812& 3.941& 3.706& 3.671& 4.125& 4.176& 4.294 & 3.871&  0.882& 0.735 &0.781 \\
            
            \midrule
            
            Correlation & 0.536 & 0.472 & 0.575 & / & -0.002 & -0.061 & 0.213 & / & 0.165 & 0.340 & 0.118 & / & / & /& /\\
            
            \bottomrule
        \end{tabular}
    }
\end{table}

\subsubsection{LLM's Effectiveness in Summarizing Large Modules.} In this section, we aim to check if our findings in RQ2, which is concluded based on the scores for summaries of small modules, are still correct for larger ones based on the marks from GPT agents. 

According to the previous result, the capability of the three GPT evaluators is similar: They are fair in evaluating the summaries except that they particularly prefer summaries generated by GPT-4. However, as some of the modules are very large, they may not fit into the context window size of the agents. For GPT-F, GPT-R, and GPT-S, 7, 3, and 1 out of 50 modules fail to be evaluated by the agents, respectively. 
Since the capability of the three agents for evaluating summaries is similar, while GPT-S can evaluate the most number of modules, we choose it as the evaluator. 

\begin{table}[!t]
    \footnotesize
    \tabcolsep=8pt
    \caption{Effectiveness of GPT-4 in evaluating the quality of summaries of large module units.}
    \label{table:large-module-scores}
    \centering
    \begin{tabular}{lllllllllllll}
    
            \toprule
            
            & \multicolumn{3}{c}{FMCS}& \multicolumn{3}{c}{RMCS}& \multicolumn{3}{c}{HMCS}& \multicolumn{3}{c}{AVG}\\ 
            
            \cmidrule(lr){2-4}\cmidrule(lr){5-7}\cmidrule(lr){8-10}\cmidrule(lr){11-13}& S-H& S-G& L-G& S-H& S-G& L-G& S-H& S-G& L-G & S-H& S-G&L-G \\
            
            \midrule
            
            GPT-4 & 2.706& 4.162 & 4.176  & 2.647& 3.882 & 3.881 & 3.071& 4.460 & 4.270 & 2.808&  4.168&4.109\\
            
            CodeLlama & 3.000&  2.213& 1.556 & 3.293& 2.893&  2.824 & 3.600&  4.431& 3.934  & 3.298& 3.149&2.771 \\
            
            CodeGemma & 2.565& 2.385& 1.351& 3.671& 3.820& 2.492& 3.871& 4.199& 3.512& 3.369& 3.468&2.452\\
            \bottomrule
        \end{tabular}
\end{table}

We report the result of GPT-S on the 17 small modules (i.e., the modules that are evaluated by humans) and the 33 large modules (i.e., the modules that are not evaluated by humans) in Table~\ref{table:large-module-scores}. S-H refers to the results of the 17 small modules marked by humans, S-G refers to the results of 17 small modules marked by the GPT agent, and L-G refers to the results of 33 large modules marked by the GPT agent.

Based on our previous findings, though the marks generated by LLMs may not be suitable for comparing the performance of different LLMs, the marks within the same LLM are highly correlated to humans. Therefore, these scores can serve as a metric for evaluating the performance changes of each summarization strategy when switching from small modules to large modules.

Firstly, we can find that for each of the LLMs, the average scores associated with larger modules are consistently lower than those for smaller modules. Furthermore, among the LLMs evaluated, GPT-4 exhibits the smallest decrease in performance, with a drop of only 1.42\%. In contrast, CodeLlama and CodeGemma show significantly larger declines at 12.00\% and 29.30\%, respectively. This considerable difference in performance drop can likely be attributed to the smaller context window sizes of CodeLlama and CodeGemma compared to the larger window size used by GPT-4.

Though the absolute performance of each strategy has dropped, their relative performance remains the same as our findings in RQ2.
For all three LLMs, the HMCS strategy can produce a better module-level summary than the FMCS and RMCS strategies ($p = 0.002$).
As for FMCS and RMCS, for GPT-4, the average performance of FMCS and RMCS strategies remains similar. For CodeLlama and CodeGemma, the FMCS strategy performs worse than the RMCS strategy where a possible reason is that they may generate some meaningless summaries.

\finding{The HMCS strategy maintains the leading position in generating summaries for large modules, consistent with its leading position in summarizing small modules.}

\summary{Combining the results from this RQ, we find that GPT-4 is effective as an evaluator for file-level summaries because 1) its scores highly correlate with those of humans, and 2) the essential qualitative conclusions drawn from human scores are also valid for GPT-4's scores. For module-level summaries, GPT-4 still shows a high correlation with human evaluations, though it may exhibit a preference for summaries generated by itself. Based on GPT-4's evaluation of the 33 large modules, our key findings and recommendations in RQ2 can be applied to both small and large modules.}

\section{Threats to Validity}
\label{sec:threats_to_validity}

\emph{$\bullet$ Internal Threats.}

The threat to internal validity lies in the implementation of the summarization strategies, i.e., file/module code reduction and file/module hierarchical summarization. 

In code reduction strategies, ``key information'' is not strictly defined. To retain as much semantic information as possible, we only remove method code bodies while preserving other content in the file, including \texttt{package} statement, \texttt{import} statement, \texttt{class} signature, global \texttt{field}, and \texttt{method} signature. Whether all of this information qualifies as ``key information'' remains unknown. In future work, we will explore the impact of each type of information on higher-level code summarization. 

The selection of 200 samples for our study is based on two main considerations. First, statistical power analysis showed that a sample size of 200 is sufficient to achieve reliable results. This sample size ensures that the study has enough statistical power to detect meaningful effects. Second, the manual evaluation of each sample involves significant human resources and time. Given the complexity of the task and the need for expert analysis, increasing the sample size beyond 200 would require an impractical amount of time and labor.

In hierarchical summarization strategies, for file-level code summarization, we only combine multiple methods into a method community based on call relationships. Although call relationships are the most straightforward and easy to understand, we believe there are other combinations (e.g., based on semantic relevance) worth exploring in future work. 

Furthermore, the issue of data contamination in the summaries is minimal. We assessed the quality of comments and documentation within the codebase, which are crucial for training summarization models. Our analysis found that less than 10\% of the functions had complete comments, and a significant 70\% of the projects lacked comprehensive API documentation or file-level comments. The absence of module-level comments was even more prevalent, and the existing comments were often outdated or inconsistent with the code implementation. This suggests that any potential contamination from the summaries generated is unlikely to have a significant impact on the overall results.
For module-level code summarization, we produce module summaries by performing text summarization on the text of each file-level summary within the module. The implementation of this strategy is rudimentary and does not consider whether files should be grouped in a manner similar to method communities, which we leave for future work.

\noindent\emph{$\bullet$ External Threats.}
Our study is currently only conducted on Java programs and may require modifications for extending to other programming languages. 

Theoretically, the three ACS strategies for commenting high-level code units explored in this study can be applied to any programming language, although different languages may have different hierarchical partitioning methods due to distinct coding conventions and styles.  
For example, in Python, a code file may consist of simple script statements without strict methods, thus requiring the design of new code reduction and hierarchical summarization strategies. 
We plan to validate the findings on other programming languages in future work.

\section{Conclusion}
\label{sec:conclusion}
In this paper, we conduct a detailed investigation into the summarization of higher-level code units based on LLMs. 
We explore and compare different strategies for summarizing file-level and module-level code units. 
We make several key findings in our experiments on file-level and module-level code unit summarization. 
For example, for file-level summarization, the optimal strategy is to input the full code into the LLM since it yields the highest quality summary. However, if budget constraints exist, using the reduced code strategy with a commercial model like GPT-4 can reduce token usage by approximately 74.65\% with only a 23.47\% reduction in summary quality. Meanwhile, the two hierarchical strategies perform poorly in file-level summarization by having the highest cost and producing the worst summaries.
For module-level summarization, the hierarchical strategy, in contrast, becomes the best choice by producing better summaries with shorter context. However, considering that the hierarchical strategy would cost a lot of tokens, a possible economical alternative should be the reduced code strategy as it produces slightly better summaries than the full code strategy with a much lower cost. Moreover, code models like CodeGemma might outperform GPT-4 in module-level summarization, particularly for smaller modules, as GPT-4 tends to be "lazy" and fails to provide summaries as detailed as those generated by CodeGemma.

In addition, we explore the feasibility of using LLMs to automate the evaluation of higher-level code unit summaries. 
We find that GPT-4 is effective as an evaluator for file-level summaries since its scores are highly correlated with those of humans and the key qualitative conclusions drawn from human scores are also valid based on GPT-4's scores. For module-level summaries, GPT-4 still shows a high correlation with human evaluations. However, it may exhibit a bias toward the summaries generated by itself.
Based on GPT-4's evaluation of the 33 large modules, we confirmed that the key findings and recommendations in RQ2 are applicable to both small and large modules.
We believe that these findings will contribute to advancing the research and application of higher-level code unit summarization.

\section*{Acknowledgment}
The authors would like to thank the anonymous reviewers for their insightful comments.

\bibliographystyle{ACM-Reference-Format}
\bibliography{reference}

\end{document}